\documentclass[showpacs,preprintnumbers]{revtex4}%
\usepackage{amsfonts}
\usepackage{amsmath}
\usepackage{amssymb}
\usepackage{graphicx}
\usepackage{dcolumn}
\usepackage{hyperref}%
\begin{document}
\title{On the electromagnetic form factors \\of the proton from generalized Skyrme models}
\author{Jimmy Fortier and Luc Marleau}
\email{lmarleau@phy.ulaval.ca}
\affiliation{D\'epartement de Physique, de G\'enie Physique et d'Optique, Universit\'e
Laval, Qu\'ebec, Qu\'ebec, Canada G1K 7P4}
\date{\today}

\begin{abstract}
We compare the prediction of Skyrme-like effective Lagrangians with data for
electromagnetic form factors of proton and consider the possibility of fixing
the parameters of these higher-order Lagrangians. Our results indicate that
one or two-parameter models can lead to better agreement with the data but
more accurate determination of the effective Lagragian faces theoretical uncertainties.

\end{abstract}

\pacs{Valid PACS appear here}
\maketitle

\section{\label{sec:intro}Introduction}

The Skyrme model \cite{Skyrme1961}, despite is relative successes
\cite{Adkins1983}, can only be considered as a prototype of an effective
theory of QCD. Indeed, large $N_{c}$ analysis \cite{'tHooft1974, Witten1979}
suggests that bosonization of QCD would most likely involve an infinite number
of mesons. If this is the case, then taking the appropriate decoupling limits
(or large mass limit) for higher spin mesons leads to an all-orders Lagrangian
in derivatives of pion fields. For now however, fixing the form of the
effective Lagrangian from an exact low-energy limit of QCD seems to be out of
our reach and the alternative has been to propose simple effective Lagrangians
\cite{Jackson1985,Marleau1990,Jackson1991} and rely on a few nucleons
properties to set the parameters they depend on. From that point of view, a
rather stringent test for such models lies in whether they could accurately
describe the data for the electromagnetic form factors of nucleons for
moderate values of momentum transfer.

The electromagnetic form factors of the semiclassically quantized SU(2)
skyrmion were studied systematically by Braaten et al \cite{Braaten1986} but
this first attempt did not take into account relativistic kinematical
corrections which are important for momentum transfers $Q^{2}>1$ GeV$^{2}.$
These correction were implemented to the original Skyrme model predictions by
Holzwarth \cite{Holzwarth1996,Holzwarth2002} using the prescription of Ji
\cite{Ji1991}. Actually, Holzwarth also\ introduced a second correction to the
Skyrme model to account for vector meson effects either by introducing a
multiplicative factor to reflect the contributions of the poles coming from
these mesons, or by adding the vector mesons as dynamical degrees of freedom
in the Lagrangian. Yet, in spite of remarks and suggestions in \cite{Acus2000,
Lassila2000, Holzwarth2002}, no analysis regarding how higher-order
Lagrangians could reproduce the nucleon electromagnetic form factors or
conversely, how the form factors could help construct a more accurate
effective Lagragian.

In this work, we calculate the electromagnetic form factors for a class of
higher-order (Skyrme-like) models introduced in \cite{Marleau1990}. For
simplicity, we limit our numerical analysis to one and two-parameter models
with a proper treatment of relavistic effects, and evaluate to what extent the
experimental data of the electromagnetic form factors of proton can suggest a
form of higher-order Lagrangians or discriminate among possible candidates.
Note that neutron data are ignored here since they are plagued with large uncertainties.

\section{\label{sec:Skyrme}The Skyrme model}

Let us first introduce the Lagrangian density for the Skyrme model
\begin{equation}
\mathcal{L}_{S}=-\frac{{F_{\pi}^{2}}}{{16}}\mathrm{Tr}\left(  {L_{\mu}L^{\mu}%
}\right)  +\frac{1}{{32e^{2}}}\mathrm{Tr}\left[  {L_{\mu},L_{\nu}}\right]
^{2}\label{eq:Skyrme_lagr}%
\end{equation}
where $F_{\pi}$ is the pion decay constant, $L_{\mu}$ is the left-handed
chiral courant $L_{\mu}=U^{\dag}\partial_{\mu}U$ and the Skyrme constant $e$
is a dimensionless constant. $U$ is a $SU(2)$ field related to the pion field
$\pi$ by $U=\exp(2i\pi\cdot\tau/F_{\pi})$. The field configurations with
finite energy must satisfy the boundary condition
\begin{equation}
U(r,t)\rightarrow\mathbf{1}\hspace{1cm}\text{for}|\mathbf{r}|\rightarrow
\infty.\label{eq:finitude}%
\end{equation}
These configurations fall into topological sectors characterized by
\begin{equation}
B=\frac{1}{{2\pi^{2}}}\int{d^{3}x\det\left\{  {L_{i}^{a}}\right\}  }%
=-\frac{{\varepsilon^{ijk}}}{{48\pi^{2}}}\int{\mathrm{d}^{\mathrm{3}}%
x}\mathrm{Tr}\left(  {L_{i}[L_{j},L_{k}]}\right)  ,\label{eq:charge}%
\end{equation}
a topological invariant taking integral values.

Skyrme interpreted this topological invariant as the baryon number.
Accordingly, the lowest-energy $B=1$ sector is identified with the nucleon. In
this sector, the lowest-energy field configuration is given by the hedgehog
ansatz
\begin{equation}
U(\mathbf{r})=\exp\left[  i\mathbf{\tau}\cdot\widehat{\mathbf{r}}F(r)\right]
\label{eq:hedgehog}%
\end{equation}
where $F(r)$ satisfies the boundary conditions $F(0)=\pi$ and $F(\infty)=0$.

With the convenient change of scale, we can use ${2\sqrt{2}}/{eF_{\pi}}$ and
${F_{\pi}}/{2\sqrt{2}}$ as units of length and energy respectively and rewrite
the Lagrangian density (\ref{eq:Skyrme_lagr}) as
\begin{equation}
\mathcal{L}_{1}+\frac{1}{2}\mathcal{L}_{2}=\left(  {\ -\frac{1}{2}%
\mathrm{Tr}L_{\mu}L^{\mu}}\right)  +\frac{1}{2}\left(  {\frac{1}{{16}%
}\mathrm{Tr}f_{\mu\nu}f^{\mu\nu}}\right)  .\label{eq:Skyrme_u}%
\end{equation}
where $f_{\mu\nu}=[L_{\mu},L_{\nu}]$. A pion mass term \cite{Adkins1984} is
usually added to account for the chiral symmetry breaking observed in nature
\begin{equation}
\mathcal{L}_{\pi}=\frac{{m_{\pi}^{2}F_{\pi}^{2}}}{{8}}\left(  {\mathrm{Tr}%
U-2}\right)  .\label{eq:masse_pions}%
\end{equation}
This term serves as a regulator for the magnetic radii and form factors (see
(\ref{eq:ray_M})) of nucleon otherwise they would diverge \cite{Beg1972}.

Using the hedgehog ansatz (\ref{eq:hedgehog}), one obtain the mass of the
static skyrmion
\begin{equation}%
\begin{split}
M &  =-\int\mathrm{{d^{3}x}\mathcal{L}_{S}}\\
&  =4\pi\left(  {\frac{{F_{\pi}}}{2\sqrt{2}e}}\right)  \int\limits_{0}%
^{\infty}{drr^{2}\left\{  {\ {F^{\prime2}+2\frac{{\sin^{2}F}}{{r^{2}}}}%
+\frac{{\sin^{2}F}}{{2r^{2}}}\left[  {2F^{\prime2}+\frac{{\sin^{2}F}}{{r^{2}}%
}}\right]  +2\beta^{2}(1-\cos F)}\right\}  }.
\end{split}
\label{eq:masse_u}%
\end{equation}
where $r$ has now been rescaled and $\beta=2\sqrt{2}m_{\pi}/eF_{\pi}.$ The
stable static soliton is obtained by minimizing the mass and requires to solve
the chiral equation {\ }%
\begin{equation}
\left(  1+a\right)  \left[  {F^{\prime\prime}+2\frac{{F^{\prime}}}{r}%
-2\frac{{sc}}{{r^{2}}}}\right]  +a\left[  {F^{\prime2}\frac{{c}}{{s}}%
+\frac{{sc}}{{r^{2}}}-2\frac{{F^{\prime}}}{r}}\right]  -\beta^{2}%
s=0.\label{eq:chirale}%
\end{equation}
with the boundary conditions $F(0)=\pi$ and $F(\infty)=0$ for $B=1.$ For
simplicity here, we used $a=\sin^{2}F/r^{2}$, $s=\sin F$, $c=\cos F$.

Fluctuations around this static soliton soliton should be quantized.
Quantization of the skyrmion is usually performed with the introduction of
spin and isospin rotation matrix as a collective coordinate \cite{Adkins1983}.
The spin/isospin rotations of skyrmion takes the form
\begin{equation}
U(\mathbf{r},t)=A^{\dag}(t)U(\mathbf{r})A(t)\label{eq:quanti}%
\end{equation}
with $A(t)$ an arbitrary time-dependent SU(2) matrix. Substituting
(\ref{eq:quanti}) in (\ref{eq:Skyrme_u}), one gets
\begin{equation}
L=-M+\mathcal{I}\,\mathrm{Tr}[\partial_{t}A^{\dag}\partial_{t}A]=-M+\frac
{I(I+1)}{2\mathcal{I}}%
\end{equation}
where $M$ is defined in (\ref{eq:masse_u}) and
\begin{equation}
\mathcal{I}=\frac{{8\pi}}{3}\left(  {\frac{{2\sqrt{2}}}{{e^{3}F_{\pi}}}%
}\right)  \int\limits_{0}^{\infty}{r^{2}dr\,\sin F(r)\left[  2+\frac{\sin
F(r)}{r^{2}}+F^{\prime2}\right]  }%
\end{equation}
is the moment of inertia of the skyrmion. Here $I$ is the spin or isospin of
the nucleon. The parameters ${F_{\pi}}$ and $e$ are fixed using two
experimental input (mass of the nucleons or else). One is then able to
reproduce the static properties of baryons within a 30 \% accuracy
\cite{Adkins1983}. These methods can also be generalized to extensions of the
Skyrme model as we will show in the following section.

\section{\label{sec:all}All-orders skyrmions}

The Skyrme model is more a prototype for an low-energy pion interactions than
a full effective field theory. Higher-order terms are expected to appear in
addition to (\ref{eq:Skyrme_u}) and (\ref{eq:masse_u}) but in its most general
form, the Lagrangian would involve a increasing number of terms at each order
in pion field derivatives making the treatment practically intractable. One of
us has proposed a special class of models whose energy density, assuming the
hedgehog ansatz, is at most linear in $F^{\prime2}$ \cite{Marleau1990}. This
requirement is sufficient to determine a unique term to each order in
derivatives and turns out to have deeper geometrical meaning \cite{Manton1987}.

The static energy density coming from the Lagrangian of order $2m$ in
derivatives of the field takes the form
\begin{equation}
\mathcal{E}_{m}={a^{m-1}\left[  {3a+m(b-a)}\right]  }\label{eq:energ_Marleau}%
\end{equation}
where $a=\sin^{2}F/r^{2}$ and $b=F^{\prime2}$. Using the hedgehog ansatz, the
first two terms arise from the non-linear $\sigma$ and the Skyrme terms
\begin{equation}
\mathcal{E}_{1}=-\mathcal{L}_{1}=-\frac{1}{2}\mathrm{Tr}L_{i}L^{i}=[2a+b]
\end{equation}%
\begin{equation}
\mathcal{E}_{2}=-\mathcal{L}_{2}=-\frac{1}{16}\mathrm{Tr}f_{ij}f^{ij}=a[a+2b]
\end{equation}
while the third term leads to
\begin{equation}
\mathcal{E}_{3}=-\mathcal{L}_{3}=\frac{1}{32}\mathrm{Tr}f_{\mu\nu}%
f^{\nu\lambda}f_{\lambda}^{\ \ \mu}=3a^{2}b
\end{equation}
as for the term proposed by Jackson et al \cite{Jackson1985} to allow for the
dynamics of the $\omega$ meson in the Skyrme model.

Generalizing to all order, the static energy associated to this class of
all-order Lagrangian can be written in a simple form
\begin{equation}
\mathcal{E}=\sum\limits_{m=1}^{\infty}{h_{m}\mathcal{E}_{m}=3\chi
(a)+(b-a)\chi^{\prime}(a)}%
\end{equation}
where a specific model is characterized by a choice of the parameters $h_{m}$
or equivalently of the function $\chi(a)=\sum\limits_{m=1}^{\infty}{h_{m}%
a^{m}}$ and $\chi^{\prime}(a)=\frac{{d\chi}}{{da}}$. Yet, $\chi(x)$ is not
completely arbitrary. Requiring that a unique soliton solution exists sets
some constraints on $\chi(x)$ \cite{Jackson1991}:
\begin{equation}%
\begin{split}
\frac{d}{{dx}}\chi(x) &  \geq0\\
\frac{d}{{dx}}\left(  {\frac{{\chi(x)}}{{x^{3}}}}\right)   &  \leq0\\
\frac{d}{{dx}}\left(  {\frac{1}{{x^{2}}}\frac{d}{{dx}}\chi(x)}\right)   &
\leq0
\end{split}
\label{eq:contr_Jack}%
\end{equation}
for $x\geq0$.

The mass of the soliton, including the pion mass term, is
\[
M=4\pi\left(  {\frac{{F_{\pi}}}{{2\sqrt{2}e}}}\right)  \int\limits_{0}%
^{\infty}{r^{2}dr\left\{  {3\chi\left(  a\right)  +\left(  {b-a}\right)
\chi^{\prime}\left(  a\right)  +2\beta^{2}(1-\cos F)}\right\}  }%
\]
and leads to the generalized chiral equation
\begin{equation}
\label{eq:gen_eom}\chi^{\prime}\left(  a\right)  \left[  {F^{\prime\prime
}+2\frac{{F^{\prime}}}{r}-2\frac{{sc}}{{r^{2}}}}\right]  +a\chi^{\prime\prime
}\left(  a\right)  \left[  {F^{\prime2}\frac{{c}}{{s}}+\frac{{sc}}{{r^{2}}%
}-2\frac{{F^{\prime}}}{r}}\right]  -\beta^{2}s=0.
\end{equation}
In that context, the Skyrme Lagrangian corresponds to $\chi(a)=a+\frac{a^{2}%
}{2}$. The moment of inertia of the soliton also take a simple form%
\begin{equation}
\mathcal{I}=\frac{{8\pi}}{3}\left(  {\frac{{2\sqrt{2}}}{{e^{3}F_{\pi}}}%
}\right)  \int\limits_{0}^{\infty}{r^{4}dr\,a\left[  {2\chi^{\prime}\left(
a\right)  +(b-a)\chi^{\prime\prime}(a)}\right]  }.\label{eq:inertie_gene}%
\end{equation}

Most all-orders model depend on more than two parameters. In the next section,
we analyze the behavior of the electromagnetic form factors for a few models
in the hope that these could help fix the $h_{m}$ coefficients and obtain a
better agreement with the the experimental data in general.

\section{Electromagnetic Form Factors}

The electromagnetic form factors of the proton in the Breit frame, for
spacelike momentum transfer $q^{2}>0$, are the Fourier transforms of its
electric charge and magnetic moment densities:
\begin{equation}
G_{E}^{p}\left(  {\ -q^{2}}\right)  =\frac{1}{2}\int\limits_{0}^{\infty
}{dr\left\{  {B_{0}\left(  r\right)  +B_{1}\left(  r\right)  }\right\}
j_{0}\left(  {qr}\right)  },\label{eq:fac_E}%
\end{equation}%
\begin{equation}
G_{M}^{p}(-q^{2})=M_{N}\int\limits_{0}^{\infty}{dr\left\{  {\frac{{4r^{2}%
B_{0}(r)}}{{e^{2}F_{\pi}^{2}\mathcal{I}}}+\mathcal{I}B_{1}(r)}\right\}
\frac{{j_{1}(qr)}}{{qr}}}\label{eq:fac_M}%
\end{equation}
with
\begin{equation}
B_{0}\left(  r\right)  =\frac{{\ -2}}{\pi}\sin^{2}(F)F^{\prime}%
,\label{eq:den_bar}%
\end{equation}
and
\begin{equation}
B_{1}\left(  r\right)  =\frac{{8\pi}}{{3\mathcal{I}}}\left(  {\frac{{2\sqrt
{2}}}{{e^{3}F_{\pi}}}}\right)  u^{4}a\left[  {2\chi^{\prime}\left(  a\right)
+\left(  {b-a}\right)  \chi^{\prime\prime}\left(  a\right)  }\right]
\label{eq:den_inertie}%
\end{equation}
are the baryon density and moment of inertia density respectively. Here
$j_{n}$ is the spherical Bessel function of order $n$ and $M_{N}$ is the
nucleons mass.

Both densities (\ref{eq:den_bar}) and (\ref{eq:den_inertie}) are normalized
\begin{equation}
\int\limits_{0}^{\infty}{drB_{0}(r)=}\int\limits_{0}^{\infty}{drB_{1}(r)=}1
\end{equation}
while the electromagnetic form factors satisfy the normalization condition
\begin{equation}
G_{E}^{p}(0)=1,\hspace{1cm}G_{M}^{p}(0)=\mu_{p}%
\end{equation}
where
\begin{equation}
\mu_{p}=\frac{M_{N}}{3}\left(  \frac{r_{B}^{2}}{2\mathcal{I}}+\mathcal{I}%
\right) \label{eq:magn_mom}%
\end{equation}
is the magnetic moment of the proton with baryonic square radius
\begin{equation}
r_{B}^{2}=\frac{8}{e^{2}F_{\pi}^{2}}\int\limits_{0}^{\infty}{dr\,r^{2}%
B_{0}(r)}.\label{eq:bar_r2}%
\end{equation}

However, the definitions (\ref{eq:fac_E}) and (\ref{eq:fac_M}) only hold in
the Breit frame moving at\ velocity $v$\ with respect to the nucleon rest
frame where the chiral profile $F(u)$ is computed. A correction for this
Lorentz boost must be applied. Ji \cite{Ji1991} has proposed a simple
prescription to circumvent this difficulty:
\begin{equation}
G_{E}(q^{2})=G_{E}^{nr}\left(  {\frac{{q^{2}}}{{\gamma^{2}}}}\right)
,\label{eq:Ge_rel}%
\end{equation}%
\begin{equation}
G_{M}(q^{2})=\gamma^{-2}G_{M}^{nr}\left(  {\frac{{q^{2}}}{{\gamma^{2}}}%
}\right)  .\label{eq:Gm_rel}%
\end{equation}
where $G_{E}^{nr}$ and $G_{M}^{nr}$ are given respectively by (\ref{eq:fac_E})
and (\ref{eq:fac_M}) and $\gamma$ is the Lorentz factor
\begin{equation}
\gamma^{2}=(1-v^{2})^{-1}=1+\frac{{q^{2}}}{{4M^{2}}}\label{eq:correction}%
\end{equation}
with the nucleon mass $M$.

Unfortunately, the boost transformations (\ref{eq:Ge_rel}) and
(\ref{eq:Gm_rel}) violate the so-called superconvergence rule
\begin{equation}
q^{2}G_{E,M}(q^{2})\rightarrow0,\hspace{1cm}\text{for }q^{2}\rightarrow
\infty.\label{eq:supercon}%
\end{equation}
which is expected to hold for electromagnetic form factors. Indeed the limit
$q^{2}\rightarrow\infty$ in the Breit frame corresponds to $q^{2}=4M^{2}$ in
the rest frame and generally $G_{E,M}^{nr}\left(  {4M}^{2}\right)  $ does not vanish.

A possible approach to restore superconvergence is to relax the condition that
$M$ must take the value of the nucleon mass and instead allow $M$ to vary in
(\ref{eq:correction}) in order to get the best agreement with the data of
$G_{M}^{p}/(\mu_{p}G_{D})$ at the highest available values of $q^{2} $
\cite{Sill1993}. However for the models under study here, this procedure
turned out to be unsatisfactory. The value of $M$ ensuring superconvergence
caused the ratio of the electromagnetic form factors to become too suppressed
in the large-$q^{2}$ limit. On the other hand, the models which we propose as
candidates for the description of QCD at low energy are not expected to hold
for large values of $q^{2}.$ Since the superconvergence rule seems to be too
restrictive, it will not be applied here. We will instead promote $M$ as
parameter and adjust its value to provide a better fit of high-$q^{2}$ data.

\section{Results and discussion}

The parameters of the Skyrme model, $e$ and $F_{\pi},$ are usually set with
either one of the following two methods: (i) adjust $e$ and $F_{\pi}$ to
obtain the mass of the nucleons (939 MeV) and of the $\Delta$ resonance (1232
MeV) or (ii) set $F_{\pi}$ according to its experimental value (186 MeV) and
adjust $e$ to reproduce the nucleon-$\Delta$ mass split (295 MeV). For
comparison purposes, we adopt the second method also used in
\cite{Holzwarth1996,Holzwarth2002} and assume that the pion mass takes its
physical value.

We consider here two simple extensions of the Skyrme model that fall into the
class of models described in Section \ref{sec:all}%

\begin{align}
\text{Model A}  & \text{:\qquad}\chi_{\text{A}}(a)=a+\frac{a^{2}}{2}%
+ca^{3}\label{eq:modA}\\
\text{Model B}  & \text{:\qquad}\chi_{\text{B}}(a)=a+\frac{{c_{1}a^{3}}%
}{{(1+c_{2}a)}}\label{eq:modB}%
\end{align}
Model A adds a single term (and a single parameter $c$) of order six in
derivative of the field\ to the Skyrme model whereas Model B which depends on
two additionnal parameter has a rational form which hopefully could allow to
reproduce poles due to vector mesons. This latter model is a generalization of
a model introduced by Jackson et al \cite{Jackson1991} with $c_{1}=\frac{1}%
{3}$ and $c_{2}=\frac{2}{3}$.

Fixing the model parameters $e,c,c_{1}$ and $c_{2}$ as well as the scale
parameter $M$ is a tedious procedure which requires a few steps: First, we
choose a set of parameters ($c$ for Model A or $c_{1}$ and $c_{2}$ for Model
B) and solve the differential equation (\ref{eq:gen_eom}) for massive pions
and use the method described above (inputs are $F_{\pi}$ and nucleon-$\Delta$
mass split) to fit for the appropriate value of $e.$ In the second step, we
compute the form factors, compare them with data and adjust $M$ to minimize
the $\chi^{2}$ of the ratio of the electromagnetic form factors. The whole
procedure is repeated with different sets of parameters until we get the
configuration $e,c,M$ or $e,c_{1},c_{2},M$ with lowest $\chi^{2}.$

\begin{figure}[tbh]
\includegraphics[width=0.8\textwidth]{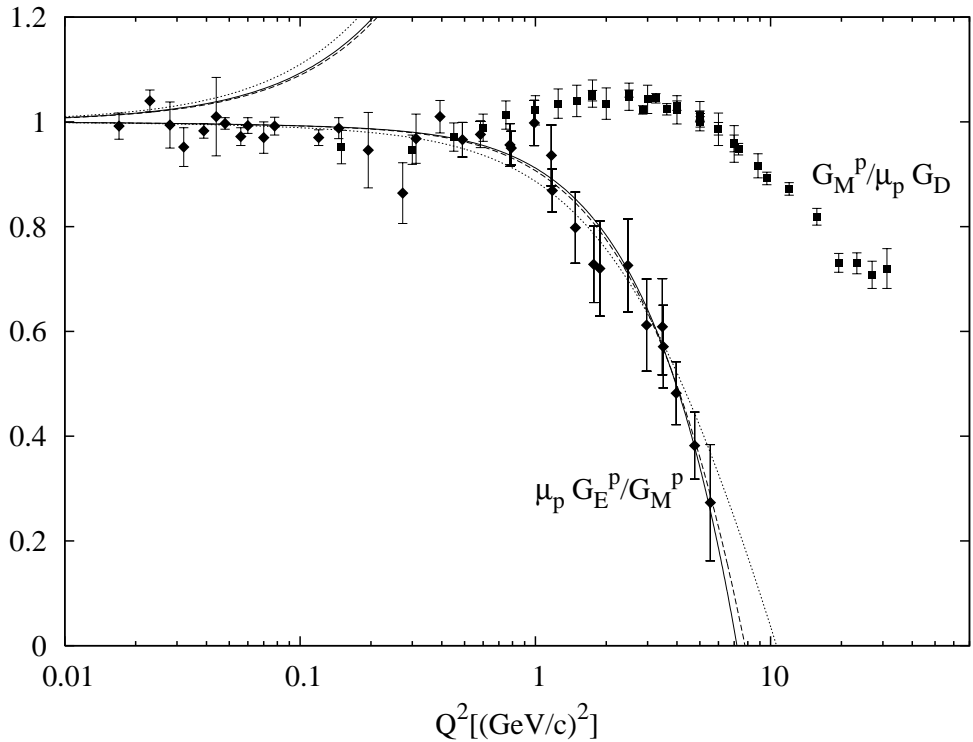}\caption{Electromagnetic
form factors of proton from Model A: $c=\frac{1}{27} $, $e=5.03$ (full lines),
Model B: $c_{1}=\frac{1}{3}$, $c_{2}=1$, $e=3.40$ (dashed lines) and Skyrme
model: $e=4.25$ (dotted lines). The data are from
\cite{Jones2000,Hohler1976,Gayou2001} (full diamonds) and
\cite{Sill1993,Andivahis1994,Walker1994} (full squares).}%
\label{fig:no_lambda}%
\end{figure}

The results for the electromagnetic form factor are presented in Fig.
\ref{fig:no_lambda}.We get $e=5.03,$ $c\simeq\frac{1}{27}$ and $M=1.20$ GeV
for Model A and $e=3.40,$ $c_{1}\simeq\frac{1}{3},$ $c_{2}\simeq1$ and
$M=1.20$ GeV for Model B. The results for the Skyrme model are also shown for
comparison ($e=4.25$ and $M=1.66$ GeV). According to Fig. \ref{fig:no_lambda},
both models present improvements over the Skyrme model, especially for the
ratio $G_{E}/\mu_{p}G_{M}^{p}$ of the form factors. Despite its simplicity,
Model A seems to overcome Model B. However both models exhibits a rapid
divergence of the magnetic form factor of proton.

In view of these results, we must conclude that although the models examined
here provide clear improvements, they cannot reproduce the data for the
magnetic form factor adequately. The problem originate to their inability to
mimic the pole coming from the vector mesons. Computing these predictions for
an arbitrary number of Skyrme model extensions being prohibitive, we resort to
another approach \cite{Holzwarth1996,Holzwarth2002}. The $\rho-$ meson effects
are incorporated by multiplying the form factors (\ref{eq:fac_E}) and
(\ref{eq:fac_M}) by
\begin{equation}
\Lambda(q^{2})=\lambda\left(  \frac{M_{\rho}^{2}}{M_{\rho}^{2}+q^{2}}\right)
+(1-\lambda)\label{eq:lamb}%
\end{equation}
with $M_{\rho}$ = 770 MeV. The electromagnetic mean square radii of proton
then become
\begin{equation}
\left\langle r^{2}\right\rangle _{E}^{p}=-\frac{6}{G(0)}\frac{dG_{E}%
^{p}(-q^{2})}{dq^{2}}\Bigm|_{q^{2}=0}=\frac{6\lambda}{M_{\rho}^{2}}+\frac
{4}{e^{2}F_{\pi}^{2}}\int\limits_{0}^{\infty}{dr\,r^{2}\left\{  {B_{0}\left(
r\right)  +B_{1}\left(  r\right)  }\right\}  },\label{eq:ray_E}%
\end{equation}%
\begin{equation}
\left\langle r^{2}\right\rangle _{M}^{p}=-\frac{6}{G(0)}\frac{dG_{E}%
^{M}(-q^{2})}{dq^{2}}\Bigm|_{q^{2}=0}=\frac{6\lambda}{M_{\rho}^{2}}+\frac
{8}{e^{2}F_{\pi}^{2}}\frac{M_{N}}{3\mu_{p}}\int\limits_{0}^{\infty}%
{dr\,r^{2}\left\{  {\frac{{4r^{2}B_{0}(r)}}{{e^{2}F_{\pi}^{2}\mathcal{I}}%
}+\mathcal{I}B_{1}(r)}\right\}  }\label{eq:ray_M}%
\end{equation}
This correction introduced an additional parameter whose allowed values go
from the purely pionic model ($\lambda=0$) to the mesonic dominance
($\lambda=1$). It should be noted that the parameter $\lambda$ does not have
any effect on the ratio of the electromagnetic form factors. A similar effect
could also be included to take into account the $\omega-$ meson effects, but
corresponding factor would involve a second $\lambda$ parameter and our
calculations indicates that this additionnal factor does not leads to
noticeable amelioration of the behavior. \begin{figure}[tbh]
\includegraphics[width=0.8\textwidth]{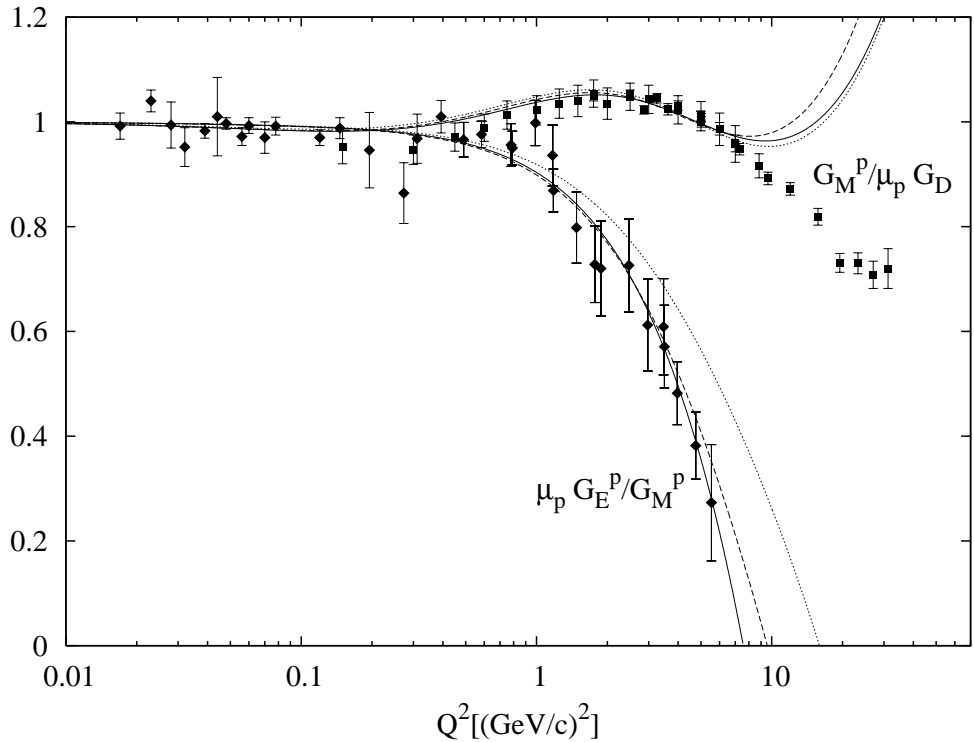}\newline%
\caption{Electromagnetic form factors of proton from Model A: $c=\frac{1}{54}
$, $e=4.74$ (full lines), Model B: $c_{1}=\frac{1}{3}$, $c_{2}=1$, $e=3.40$
(dashed lines) and Skyrme model: $e=4.25$ (dotted lines). The data are the
same as for Fig..~\ref{fig:no_lambda}.}%
\label{fig:modAB}%
\end{figure}

Figure \ref{fig:modAB} presents the electromagnetic form factors as obtained
from Model A and B with pole effects introduced through (\ref{eq:lamb}). All
the models parameters as well as the parameters $M$ and now $\lambda$ are
fitted using the procedure described above to minimize the $\chi^{2}$ of the
magnetic form factor. The best accord with data are acheived for values of the
parameters $e=4.74$ $c\simeq\frac{1}{54}$ with $M=1.31$ GeV and $\lambda=0.73$
for Model A and $e=3.40,$ $c_{1}\simeq\frac{1}{3}$ and $c_{2}\simeq1$ with
$M=1.26$ GeV and $\lambda=0.72$ for Model B. The results for the Skyrme model
are found to be constistent with those of refs.
\cite{Holzwarth1996,Holzwarth2002} with $M=1.42$ GeV and $\lambda=0.75$.
Clearly, the inclusion of vector mesons effects produce a significant
improvement $G_{M}^{p}/\mu_{p}G_{D}$ over the results of Fig.
\ref{fig:no_lambda}. Again both extensions of the Skyrme model, Model A and B,
are better at reproducing the data with Model A being slightly superior
despite its relative simplicity. Furthermore, as shown in table
\ref{tab:results}, the addition of higher orders terms have little effect on
other static properties of the proton with these choice of parameters (Note
that the magnetic moment of proton predicted by the models are all lower than
its experimental value but one has to remember that these properties are
subject to quantum corrections \cite{Meier1997}). The values of $\lambda$
indicate a strong dominance of the $\rho-$ meson as in
\cite{Holzwarth1996,Holzwarth2002}.

Our results therefore indicates that the data are best reproduced by a
higher-order effective Lagrangian in the low-energy limit of QCD .
Unfortunately, as we have noticed, the determination of a more acurate
effective Lagrangian faces theoretical uncertainties which remain to be
addressed. First, the boost prescription (\ref{eq:Ge_rel}) and
(\ref{eq:Gm_rel}), not being compatible with superconvergence, prevents us
from reproducing results for $q^{2}>10\,\text{(GeV/c)}^{2}$. An adequate boost
prescription holding account of superconvergence should allow to extend the
analysis to higher momentum tranfer. Secondly, we have only explored the
possibilities of two types of Skyrme model extensions. These could not mimick
the vector meson effects satisfactorily without the artifact of equation
(\ref{eq:lamb}). Finding a form of Skyrme-like lagrangian that would allow to
bypass the approach in (\ref{eq:lamb}) and avoid the introduction of the
$\lambda$ parameter remains a challenge.%

\begin{table}[tbp] \centering
\begin{tabular}
[c]{lccccccc}\hline
& Skyrme & \hspace{0.5cm} & Model A & \hspace{0.5cm} & Model B &
\hspace{0.5cm} & Exp.\\\hline\hline
$M$ (GeV$^{{}}$) & $1.43$ &  & $1.31$ &  & $1.26$ &  & $-$\\
$\lambda$ & $0.75$ &  & $0.73$ &  & $0.72$ &  & $-$\\
$\mu_{p}$ ($\mu_{N}$) & $1.78$ &  & $1.79$ &  & $1.80$ &  & $2.79$\\
$r_{E}^{p}$ (fm) & $0.684$ &  & $0.701$ &  & $0.706$ &  & $0.870\pm0.008$\\
$r_{M}^{p}$ (fm)\vspace{0.5cm} & $0.826$ &  & $0.841$ &  & $0.843$ &  &
$0.858\pm0.056$\\\hline
\end{tabular}
\caption{Static nucleon properties and parameters for models of FIG.\protect
\ref{fig:modAB}. The experimental data come from \protect\cite{Hagiwara2002,Dumbrajs1983}.}
\label{tab:results}
\end{table}%

\end{document}